\documentclass[letterpaper, 10 pt, conference]{ieeeconf}  
\IEEEoverridecommandlockouts
\usepackage{cite}
\usepackage{amsmath,amssymb,amsfonts}
\usepackage{booktabs}
\usepackage{algorithm}
\usepackage{algpseudocode}

\usepackage{bm} 
\usepackage{graphicx}
\usepackage{textcomp}
\usepackage{xcolor}
\usepackage{cite}
\usepackage{array}
\usepackage{multirow}
\usepackage{siunitx}     

\newcommand{\bx}{\bm{x}}
\newcommand{\bu}{\bm{u}}
\newcommand{\by}{\bm{y}}
\newcommand{\R}{\mathbb{R}}
\newcommand{\E}{\mathbb{E}}

\newcommand{\ddt}[1]{\dfrac{d #1}{d t}}

\newtheorem{problem}{Problem}
\usepackage[printonlyused]{acronym}
\acrodef{AID}{Automated Insulin Delivery}
\acrodef{CGM}{Continuous Glucose Monitor}
\acrodef{CHO}{Carbohydrate}
\acrodef{DS}{Decision Support}
\acrodef{KF}{Kalman Filter}
\acrodef{EKF}{Extended Kalman Filter}
\acrodef{AEKF}{Augmented Extended Kalman Filter}
\acrodef{EGP}{Endogenous Glucose Production}
\acrodef{MPC}{Model Predictive Control}
\acrodef{OHSU}{Oregon Health \& Science University}
\acrodef{PF}{Particle Filter}
\acrodef{RBSVGD}{Rao-Blackwellized Stein Variational Gradient Descent}
\acrodef{SVGD}{Stein Variational Gradient Descent}
\acrodef{T1D}{Type 1 Diabetes}
\acrodef{UKF}{Unscented Kalman Filter}
\title{\LARGE \bf Joint State-Parameter Inference Enhances Estimation Performance in Model-Based Digital Therapeutics for Type 1 Diabetes
}

\author{Milad Banitalebi Dehkordi, Vihangkumar V. Naik, Manas Mejari, Dario Piga, Jose Garcia-Tirado$^{\star}$
\thanks{This work has been accepted for publication at the 65th IEEE Conference on Decision and Control (CDC 2026), Honolulu, Hawaii, USA. M. B. Dehkordi, M. Mejari and D. Piga are with Dalle Molle Institute for Artificial Intelligence, IDSIA-SUPSI, Via la Santa 1, CH-6962 Lugano-Viganello, Switzerland \texttt{\{milad.banitalebi, manas.mejari, dario.piga\}\@supsi.ch}.  V. V. Naik and J. Garcia-Tirado are with Department of Diabetes, Endocrinology, Nutritional Medicine, and Metabolism, Inselspital, Bern University Hospital and University of Bern,
Switzerland and Diabetes Center Berne, Bern, Switzerland \texttt{\{vihangkumar.naik, jose.garcia\}\@unibe.ch}. $^{\star}$Corresponding author.}
}

\begin{document}
\maketitle
\thispagestyle{empty}
\pagestyle{empty}

\begin{abstract}
Blood glucose estimation is the cornerstone of model-based decision support (DS) and Automated Insulin Delivery (AID) systems. Control systems that rely on physiologic/compartmental models depend heavily on model parameterization, which is either defined using population values or personalized through the user’s data. Often, the model parameters are defined as constants. However, under real-world free-living conditions, fixed parameters can limit the accurate reconstruction and estimation of glucose levels and states. 
In this paper, we propose and discuss a recursive filtering framework for online joint state estimation and parameter identification in nonlinear, time-varying physiological models for Type 1 Diabetes (T1D). Specifically, we employ a Rao-Blackwellized Stein Variational Gradient Descent (RBSVGD) filter to compute the joint posterior distributions of model states and parameters. The proposed approach is applied to the Hovorka glucose-insulin model and validated using data generated by the \ac{OHSU} simulator across 20 virtual patients. We perform a comparative analysis against: (i) a standard Extended Kalman Filter (EKF) with fixed model parameters, and (ii) an Augmented Extended Kalman Filter (AEKF) for joint state-parameter estimation. The results demonstrate that the proposed RBSVGD-based framework outperforms both EKF and AEKF approaches not only in terms of the accuracy of glucose estimation, but also in terms of estimated model parameters.
\end{abstract}


\section{Introduction}

Physiological/compartmental models lie at the core of \ac{AID} and \ac{DS} systems for \ac{T1D} management. In practice, these models, which contain multiple ordinary differential equations that account for the user's full metabolic state, are typically parameterized using population-level values. Under these conditions, the accuracy and reliability of state estimation and prediction may be compromised, potentially impacting decision-making. 
Parameter adaptation has been shown to be essential for personalized prediction, estimation and closed-loop control in \ac{AID} systems, as static population models cannot capture intra-and inter-patient variability, \ac{CGM} sensor drift, or context-dependent changes in insulin sensitivity due to exercise, stress, illness, or circadian rhythms~\cite{Eberle_modelAdatation_2012, UKF_paramest_Cinar_2018}. This renders the physiological models, \emph{e.g.} the Hovorka model,  of the glucose--insulin system in \ac{T1D} to be inherently nonlinear, and governed by physiological parameters
that vary continuously both within and across individuals

 These characteristics expose fundamental limitations in classical
filters employed in \ac{AID} and \ac{DS}, which rely on linear fixed-parameter state-space models combined with the \ac{EKF} for denoising \ac{CGM} sensor data,  and short-term glucose predictions~\cite{Knobbe_EKF_CGM_2005, wang_EKF_2014}. Such approaches provide limited personalization, as model parameters are typically assumed constant. To address this limitation, joint state and parameter estimation frameworks have been proposed. In~\cite{Sala-Mira_2021}, an \ac{AEKF} with parameters embedded in an augmented
state vector is evaluated for joint state-parameter estimation on the Hovorka model. However, a fundamental limitation of the \ac{AEKF}  is the propagation of a \emph{Gaussian}
belief over the joint state-parameter posterior. This  unimodal approximation
is unable to capture the multimodal uncertainty that arises in the presence of unannounced meals, exercise, or rapid changes in insulin sensitivity. The \ac{UKF}~\cite{julier2004unscented, UKF_paramest_Cinar_2018,Eberle_modelAdatation_2012} replaces linearization with
a deterministic sigma-point transform, however, the posterior remains restricted to the Gaussian family and thus cannot represent non-Gaussian features of the parameter distribution.
The \ac{PF} ~\cite{Sontakke_PF_MealDetection_Estimation_2025, Wang_PF_AdaptiveMPC_2021} removes the
Gaussian restriction by representing the posterior as a weighted particle
ensemble, but its stochastic proposal mechanism renders parameter updates
sample-inefficient and requires careful selection of proposal distributions.

In this work, we employ the \ac{RBSVGD} filter \cite{milad2026rbsvgd} which  addresses each of these deficiencies. Rao--Blackwellization analytically marginalizes
 conditionally linear-Gaussian substructure, reducing the effective
dimension of the sampling-based Monte Carlo problem. 
The parameter posterior is 
updated via \acf{SVGD}, a deterministic, gradient-driven update that
iteratively steers an ensemble of parameter particles toward high-probability
regions of the posterior. Unlike the \ac{PF}, this deterministic update
achieves faster empirical convergence without requiring problem-specific
proposal design or large number of particles. Unlike the \ac{EKF}, \ac{AEKF},
and \ac{UKF}, it places no parametric restriction on the shape of the
parameter posterior, naturally representing multimodal or skewed
distributions that arise under realistic \ac{T1D} conditions. The resulting
filter is fully recursive and operates in real time, requiring no offline
system identification  as in \cite{sun2023adaptive}. 
These properties translate directly to clinically meaningful advantages for
\ac{T1D} management. Estimating a subset of Hovorka model parameters online
within a non-Gaussian posterior allows the filter to capture intra-day
insulin-sensitivity fluctuations and patient--specific parameters,
enabling a continuously personalized internal model that adapts to patients' physical activity, and inter-patient variability without manual
recalibration. The in silico data from the \ac{OHSU} simulator is used for more realistic experimental validation~\cite{OHSU_simulator_2019}. Overall, the proposed RBSVGD filter demonstrates superior filtering and parameter estimation performance compared to the EKF and AEKF.

\textit{Notations:} Let $\R^n$ be the set of real vectors with dimension $n$. The $\bx_{1:k}$ denotes the set of vectors $\{ \bx_1, \ldots, \bx_k\}$ from time index $1$ to $k$. We denote by $\bx_{k|t}$ the value of $\bx$  computed at time $k$, given the measurements up to time $t$. Let $w \sim \mathcal{N}(\mu,P)$ denote a random variable $w$ having a Gaussian distribution with mean $\mu$ and covariance $P$. The notation $\delta_{\bar{\theta}}(\theta)$ denotes the Dirac delta function of a random variable $\theta$ centered at $\bar{\theta}$. The expected value and the covariance of a random variable are denoted as $\E[x]$ and $\rm cov[x]$, respectively.

\section{Problem Statement}\label{sec:problem_set}
We adopt the Hovorka glucose-insulin model~\cite{Hovorka2004} (\emph{cf.} Appendix), which has become a widely accepted benchmark for model-based glucose control research.  
The discretization of the Hovorka glucose-insulin model is expressed as the following nonlinear parameter-varying representation:
\begin{subequations}\label{eq:system}
    \begin{align}
    \bx_{k+1} &= f(\bx_k, \bu_k, \theta^{\star}_k) + w_k,  \label{eq:system_a} \\
    \by_k & = g(\bx_k, \theta^{\star}_k) + v_k. \label{eq:system_b}
\end{align}
\end{subequations}

The model state $\bx_k \in \R^{10}$ is defined as:
\begin{align}
  \bx_k = \bigl[
    Q_{1,k},\; Q_{2,k},\; S_{1,k},\; S_{2,k},\; I_k,\;\notag\\
    x_{1,k},\; x_{2,k},\; x_{3,k},\; D_{1,k},\; D_{2,k}
  \bigr]^{\!\top},
\end{align}

\noindent
where $Q_{1,k}$~(mmol) and $Q_{2,k}$~(mmol) are the glucose masses in the
accessible (plasma) and non-accessible (interstitial) compartments,
respectively; $S_{1,k}$~(mU) and $S_{2,k}$~(mU) are two successive insulin
absorption compartments modeling the subcutaneous depot dynamics; $I_k$~(mU/L)
is the plasma insulin concentration; $x_{1,k}$~(min$^{-1}$), $x_{2,k}$~(min$^{-1}$),
and $x_{3,k}$~(min$^{-1}$) are remote insulin effect variables driving,
respectively, glucose distribution, glucose disposal, and \ac{EGP} suppression;
and $D_{1,k}$~(mmol) and $D_{2,k}$~(mmol) are the two gut carbohydrate
absorption compartments.
The input vector $\bu_{k} = [u_{1,k},\, u_{2,k}]^{\top}$ consists of: $u_{1,k}$ (mU/min): the exogenous subcutaneous insulin infusion rate; and
$u_{2,k}$ (mg/min): the rate of carbohydrate ingestion, representing meal disturbances. The  model parameters are defined as:
\begin{align}\label{eq:hovorka_fixed_params}
  \theta^\star = \bigl[
    k_{12},\; k_{a_{1}},\; k_{a_{2}},\; k_{a_{3}},\; k_{e},\;V_G,\; V_I,\;  A_G,\; \notag\\\;S_{I,1},\; S_{I,2},\; S_{I,3}, F_{01}, EGP_0\; \tau_D,\; \tau_S
  \bigr]^{\!\top} \in \R^{15},
\end{align}
\noindent
where $k_{12}$ (min$^{-1}$) is transfer rate; $k_{a_{1}}$, $k_{a_{2}}$, $k_{a_{3}}$ (min$^{-1}$) are deactivation rates; $k_{e}$ (min$^{-1}$) is Insulin elimination from plasma; $V_G,~V_I$ (L kg$^{-1}$) are the volumes of Glucose distribution and insulin distribution respectively; $A_G$ is \ac{CHO} bioavailability; $S_{I,1},\; S_{I,2},\; S_{I,3}$ (min$^{-1}$/mU L$^{-1}$) are insulin sensitivities of distribution, disposal and \ac{EGP}; $F_{01}$ (mmol kg$^{-1}$ min$^{-1}$) is non-insulin-dependent glucose flux; $EGP_0$ (mmol kg$^{-1}$ min$^{-1}$) is \ac{EGP} extrapolated to zero insulin; $ \tau_D,\; \tau_S$ (min) are rates of absorptions for \ac{CHO} and Insulin respectively. 

The model output $\bm{y}_k \in \R $ is the  plasma glucose concentration, given by  $G_k = Q_{1,k}/V_G$. The process and measurement noises are modeled as zero-mean Gaussian\footnote{Note that the non--Gaussian (process and measurement) noises are absorbed by time--varying parameters $\theta^{\star}_k$.}, \emph{i.e.}, $w_k \sim \mathcal{N}(0,Q), v_k \sim \mathcal{N}(0, R)$ with covariances $Q, R \succeq 0$.


In this work, we address the joint estimation of the state $\bx_k$ and a subset 
$\theta_{k} \subset \theta^{\star}_k$, 
from noisy blood-glucose measurements $\by_k$ and insulin and meal records $\bu_k$. Within a Bayesian framework, the objective is to infer the joint posterior $p(\bx_k, \theta_k | \by_{0:k}, \bu_{0:k})$ over the states and parameters given the measurements up to time $k$. To simplify the notation, we omit the dependence of the posterior distribution $p(\bx_k,\theta_k|\by_{1:k}, \bu_{1:k})$ on the input sequence $\bu_{1:k}$ in the rest of the paper. 
The problem considered in this paper is formalized as follows. 
\begin{problem}\label{prob}
Given noisy blood-glucose measurements $\bm{y}_k$ and the inputs $\bu_k$ at each sampling time $k$, estimate recursively the joint posterior distribution $p(\bx_k,\theta_k|\by_{1:k}) $ of states $\bm{x}_k$ and time-varying physiological parameters $\theta_{k}$.  
Based on this posterior, compute point estimates, \emph{e.g.} maximum a posteriori (MAP) or posterior mean, of $\bm{x}_k$ and $\theta_k$. \hfill $\blacksquare$
\end{problem}




Problem~\ref{prob} is addressed in the next section by the \ac{RBSVGD} filter,
which exploits the factorization of the joint posterior into a conditional
state distribution and a marginal parameter distribution to decouple---and
solve efficiently---the two subproblems of state filtering and parameter
tracking.

\section{Rao-Blackwellized SVGD filter}


Let us express the joint posterior  $p(\bx_k, \theta_k | \by_{1:k})$ as 
\begin{equation}\label{eq:target_posterior}
p(\bx_k, \theta_k | \by_{1:k} ) = p(\bx_k| \theta_k, \by_{1:k}) \, p(\theta_k | \by_{1:k}). 
\end{equation}
In our RBSVGD algorithm, an approximation of the  conditional distribution $p(\bx_k| \theta_k, \by_{1:k})$  is computed \emph{analytically} using an \emph{Extended Kalman Filter} (EKF), while the marginal posterior $p(\theta_k | \by_{1:k})$ is approximated using  a  \emph{Stein-Variational Gradient Descent} (SVGD) filter~\cite{liu2016stein}. 

In particular, the  marginal posterior is approximated by running an SVGD--based filter as,
\begin{align}\label{eq:SVGD_post}
  p(\theta_k | \by_{1:k}) \approx \frac{1}{N}\sum \limits_{i=1}^{N} \delta_{\theta^{(i)}_k}(\theta_k),  
\end{align}
where $\{ \theta^{(i)}_k\}_{i=1}^{N}$ are $N$ sampled particles (each representing a different model). 
For each particle $\theta^{(i)}_k$,  the conditional state posterior is approximated as a Gaussian distribution through  EKF, i.e.,  
\begin{align}\label{eq:EKF_post}
   p(\bx_k| \theta^{(i)}_k, \by_{1:k}) \approx  \mathcal{N}(\bx_k; \mu_{k|k}, P_{k|k}), 
\end{align}
where  $\mu_{k|k} $ and $P_{k|k} \succeq 0 $ denote the mean and  covariance at time $k$ computed from the EKF iterations using the measurements $\by_{1:k}$ up to time $k$. From \eqref{eq:SVGD_post} and \eqref{eq:EKF_post}, the target  posterior in \eqref{eq:target_posterior} is then approximated as a Gaussian mixture,
\begin{align}\label{eq:approx_posterior}
p(\bx_k, \theta_k | \by_{1:k} ) \approx  \frac{1}{N} \sum_{i=1}^{N} \mathcal{N}(\bx_k; \mu_{k|k}^{(i)}, P_{k|k}^{(i)})  \delta_{\theta_k^{(i)}}(\theta_k).
\end{align}

This approach enables efficient posterior approximation by restricting SVGD sampling to the parameter space $\theta_k \in \Theta \subset \mathbb{R}^{n_\theta}$, thus reducing computational complexity compared to sampling over the joint space $(\bx_k,\theta_k)$.
Unlike the \emph{Augmented Extended Kalman Filter} (AEKF), which assumes a Gaussian approximation of the joint state-parameter posterior, the proposed method does not impose a parametric form on $p(\theta_k | \by_{1:k})$. This allows the representation of non-Gaussian and potentially multi-modal parameter distributions.

We now describe the  RBSVGD filter proposed in~\cite{milad2026rbsvgd}. We begin by recalling the EKF recursions used to compute the conditional state posterior associated with each particle $\theta_k^{(i)}$. Subsequently, we describe how the particles $\{\theta_k^{(i)}\}_{i=1}^N$ are updated via SVGD to approximate the marginal posterior $p(\theta_k \mid \by_{1:k})$. 

\subsection{EKF for the state posterior $p(\bx_k|\theta^{(i)}_k,\by_{1:k})$}

For each particle $\theta^{(i)}_k$ (namely, model parameters), an EKF is employed to estimate the state $\bx_k$. Let $\mu^{(i)}_{k|k} := \E{[\bx_k|\by_{1:k}, \theta_k]}$  and $P^{(i)}_{k|k} := \mathrm{cov}[\bx_k |\by_{1:k}, \theta^{(i)}_k]$  denote the mean and the covariance of the state at time $k$, corresponding to the $i$-th particle $\theta^{i}$. 

The EKF procedure  at time $k$ is summarized in \textbf{Algorithm~\ref{alg:EKF}}, which requires the particle  $\theta^{(i)}_k$, along with initial state mean $\bar{\mu}_0$ and  covariance $\bar{P}_{0|0}$. 

At the initialization Step~\ref{alg:s0}, the EKF state mean $\mu_{k-1|k-1}$ and covariance $P_{k-1|k-1}$ are initialized to $\bar{\mu}_0$ and $\bar{P}_{0|0}$ respectively. In the prediction Step \ref{algo:ekf_pred}, the predicted mean and state covariance  $\mu_{k|k-1}^{(i)}$, $P_{k|k-1}^{(i)}$  are computed based on linearizing the model \eqref{eq:system}, where 
  $F_{k-1}^{(i)} \!=\! \frac{\partial{f}}{\partial{\bx}}$ in \eqref{eq:ekf_prediction} is the Jacobian of the model dynamics $f$ in \eqref{eq:system_a}, evaluated at the previous mean $\mu_{k-1|k-1}^{(i)}$ and $\theta_{k-1}^{(i)}$. Step~\ref{algo:ekf_gain} computes the innovation covariance $S_k^{(i)}$ based on the Jacobian $H_k^{(i)} \!=\! \frac{\partial{g}}{\partial{\bx}}$ of the measurement model $g$ in \eqref{eq:system_b},  evaluated at $\mu_{k|k-1}^{(i)}$ and $\theta_{t-1}^{(i)}$. The Kalman gain $K_k^{(i)}$ is computed as in \eqref{eq:kalman_gain} based on the innovation covariance.

Finally, in the EKF update Step~\ref{algo:ekf_update},  
the state mean  and covariance   are updated according to \eqref{eq:ekf_update}, using  the glucose measurement $\by_k$, and the Kalman gain $K_k^{(i)}$ which balances model predictions and measurements.    
The algorithm returns the estimated mean and covariances  $\{\mu_{k|k}^{(i)}, P_{k|k}^{(i)}\}$.

\begin{algorithm}[t!]
\caption{EKF for state estimation}
\begin{algorithmic}[1]
\Require  Particle $\theta^{(i)}_k$,  initial  state mean $\bar{\mu}_{0}$; initial state covariance $\bar{P}_{0}$.
\State \label{algo:s0} \textbf{Initialization:}\label{alg:s0}
\Statex \quad Set $\mu_{k-1|k-1}^{(i)} = \bar{\mu}_0,\, P_{k-1|k-1}^{(i)} = \bar{P}_{0}$   
\State \label{algo:ekf_pred} \textbf{EKF prediction:}
\begin{subequations}
\label{eq:ekf_prediction}
\begin{align}
\mu_{k|k-1}^{(i)} &= f\big(\mu_{k-1|k-1}^{(i)}, \bu_{k-1}, \theta_{k-1}^{(i)}\big), \\
P_{k|k-1}^{(i)} &= F_{k-1}^{(i)} P_{k-1|k-1}^{(i)} (F_{k-1}^{(i)})^\top + Q(\theta_{k-1}^{(i)}),
\end{align}
\end{subequations}
 \State \label{algo:ekf_gain}  \textbf{Compute the Kalman gain:} 
 \begin{subequations}
    \begin{align}
S_k^{(i)} &= H_k^{(i)} P_{k|k-1}^{(i)} (H_k^{(i)})^\top + R(\theta_{k-1}^{(i)}), \label{eq:innov} \\
K_k^{(i)} &= P_{k|k-1}^{(i)} (H_k^{(i)})^\top \big( S_k^{(i)} \big)^{-1} \label{eq:kalman_gain}
\end{align}  
 \end{subequations}

\State \label{algo:ekf_update}  \textbf{EKF  update:}
        \begin{subequations}
\label{eq:ekf_update}
\begin{align}
\mu_{k|k}^{(i)} &= \mu_{k|k-1}^{(i)} + K_k^{(i)} \big( \by_t - h(\mu_{k|k-1}^{(i)}, \theta_{k-1}^{(i)}) \big), \\
P_{k|k}^{(i)} &= \big( I - K_k^{(i)} H_k^{(i)} \big) P_{k|k-1}^{(i)}.
\end{align}
\end{subequations}
\Statex \Return Updated mean and covariance $\{\mu_{k|k}^{(i)}, P_{k|k}^{(i)}\}$
\end{algorithmic}
\label{alg:EKF}
\end{algorithm}


\subsection{SVGD for the parameter posterior $p(\theta_k|\by_{1:k})$ 
}
To approximate the marginal posterior $p(\theta_k \mid \by_{1:k})$ over the parameters, we employ a deterministic Stein Variational Gradient Descent (SVGD)~\cite{liu2016stein} framework. In SVGD, particles $\{\theta^{(i)}_k\}_{i=1}^{N}$ are updated via gradient-based transformations that minimize the Kullback-Leibler (KL) divergence to the target posterior. Specifically, each particle at time $k$ is iteratively updated as follows,
\begin{equation}
\label{eq:svgd_update}
\theta^{(i)}_{k,m} \leftarrow \theta^{(i)}_{k,m-1} + \epsilon \, \hat{\phi}^\ast(\theta^{(i)}_{k,m-1}),
\end{equation}
where $\theta^{(i)}_{k,m}$ denotes the $i$-th particle at iteration $m$, at time $k$, and $\epsilon >0$ is a step size. This corresponds to a deterministic parameter transition model $p
(\theta_k|\theta_{k-1})$.

The term $\hat{\phi}^{\ast}(\cdot)$ denotes the empirical \emph{perturbation direction}.  It can be proved that the optimal $\hat{\phi}^{\ast}(\cdot)$  is the direction of steepest descent on the KL divergence
between proposal and target posterior $p(\theta_k|\by_{1:k})$  \cite{milad2026rbsvgd}. In other words, the particles evolving according to \eqref{eq:svgd_update} minimize the KL divergence at each iteration and the empirical posterior of the particles converges to the target posterior.

The perturbation direction in \eqref{eq:svgd_update} is computed based on the particles $\theta^{(i)}$ and the log-likelihood $\log p(\by|\theta)$ as follows (The subscript time index $k$ and iteration $m$ are dropped for notational simplicity):
\begin{align}
\hat{\phi}^\ast(\theta) = 
\frac{1}{N} \sum_{i=1}^N &
\big[\kappa(\theta^{(i)},\theta)\,\nabla_{\!\theta^{(i)}} \log [p(\by|\theta^{(i)})p(\theta^{(i)}) ]\notag\\
&\quad  \quad  + \nabla_{\!\theta^{(i)}} \kappa(\theta^{(i)},\theta)\big],
\label{eq:svgd_empirical}
\end{align}
 where $\kappa(\theta',\theta)$ is a positive-definite kernel chosen by the user, \emph{e.g.}, the radial basis function $\kappa(\theta', \theta) = \exp\!\big(\frac{-1}{h}\|\theta' \!-\! \theta\|^2\big)$, and $\nabla_{\theta^{'}}\log [p(\by|\theta')p(\theta'|\by)]$ is the gradient of the unnormalized log-posterior evaluated at  $\theta^{'}$.
Note that the log-likelihood $\log p(\by|\theta')$ can be computed from the EKF as 
\begin{align}\label{eq:likelihood}
\log p(\by_k|\by_{1:k-1}, \theta') = \frac{-1}{2} \left(r^{\top}_k S^{-1}_kr_k + \log \det S_k \right),
\end{align}
where $r_k := \by_k - h(\mu_{k|k-1}, \theta_{k-1})$ is the residual and $S_k$ is the innovation covariance in \eqref{eq:innov}.

The perturbation direction \eqref{eq:svgd_empirical}  balances between exploration of the particles and their concentration in high-probability regions.

In summary, for iterations $m=1,2,\ldots, M$, each particle  $\theta^{(i)}_k$ at time $k$ is updated according to \eqref{eq:svgd_update}. The obtained particles $\{\theta^{(i)}_k \}_{i=1}^{N}$ then approximate the marginal posterior as given in \eqref{eq:SVGD_post}. The particle update procedure at time step $k$ is summarized in \textbf{Algorithm~\ref{alg:SVGD}}. The algorithm requires an initial set of particles, a user-chosen kernel function $\kappa(\cdot,\cdot)$, and the SVGD step size $\epsilon$. It returns updated particles $\{\theta_{k}^{(i)}\}_{i=1}^N$ at time $k$, after $M$ number of iterations.

\begin{algorithm}[t!]
\caption{SVGD for parameter estimation}
\begin{algorithmic}[1]
\Require Set of initial particles $\{\theta^{(i)}_{0,k} \}_{i=1}^{N}$; log-likelihood $p(\by_{k}|\by_{1:k-1}\theta^{(i)})$, kernel function $\kappa(\cdot,\cdot)$, step size $\epsilon$, number of iterations $M$.
\State \textbf{for} {$m = 1,2,\ldots, M$ }
\begin{align*}
\theta^{(i)}_{k,m} \leftarrow \theta^{(i)}_{k,m-1} + \epsilon \, \hat{\phi}^\ast(\theta^{(i)}_{k,m-1}), 
\end{align*}
where $\hat{\phi}^\ast(\cdot)$ as in \eqref{eq:svgd_empirical}.
\Statex \textbf{end for} 
\State $\theta^{(i)}_k \leftarrow \theta^{(i)}_{k,M}, \forall i=1,\ldots, N.$
\State \Return $\{\theta_{k}^{(i)}\}_{i=1}^N$
\end{algorithmic}
\label{alg:SVGD}
\end{algorithm}

\subsection{RBSVGD filter for joint state parameter computation}

By combining the EKF \textbf{Algorithm~\ref{alg:EKF}} and SVGD \textbf{Algorithm~\ref{alg:SVGD}}, we now present the proposed RBSVGD algorithm to compute the joint state and parameter posterior. The RBSVGD procedure at time instance $k$  is detailed in \textbf{Algorithm~\ref{alg:RBSVGD}}, which returns the set $\{\mu_{k|k}^{(i)}, P_{k|k}^{(i)}, \theta_k^{(i)}\}_{i=1}^N$ corresponding to the state means, covariances, and parameter samples for $N$ particles. This set is used to construct the approximate joint posterior as given in \eqref{eq:approx_posterior}, as well as to compute point estimates (\emph{e.g.} MAP or posterior mean) of the states and the parameters, thus solving Problem~\ref{prob}.    Note that the algorithm is fully recursive, and it can be run online as the new measurements $\by_k$ are available at each sampling instance.  For more details of the RBSVGD filter, the interested reader is referred to the report~\cite{milad2026rbsvgd}.

\begin{algorithm}[h!]
\caption{Rao--Blackwellized Stein Variational Gradient Descent Filter~\cite{milad2026rbsvgd}}
\begin{algorithmic}[1]
\Require Measurements $\by_k$, state mean and covariance $\{\mu_{k-1|k-1}^{(i)}, P_{k-1|k-1}^{(i)}\}_{i=1}^N$, previous particles $\{\theta_{k-1}^{(i)}\}_{i=1}^N$, RBF kernel $\kappa(\theta,\theta')$, SVGD step size $\epsilon >0$
\For {$i = 1$ to $N$}
  \State \textbf{EKF Predict:} $\mu_{k|k-1}^{(i)}$, $S_{k}^{(i)}$ using EKF \textbf{Algorithm~\ref{alg:EKF}}.
 \State \textbf{EKF Likelihood: as given in \eqref{eq:likelihood}.}
  \State \textbf{EKF Update:} $\mu_{k|k}^{(i)}$, $P_{k|k}^{(i)}$ using EKF \textbf{Algorithm~\ref{alg:EKF}}.
 \State \textbf{Particle Update:} 
  $\theta_{k}^{(i)}$ using SVGD \textbf{Algorithm~\ref{alg:SVGD}}.
\EndFor
\State \Return $\{\mu_{k|k}^{(i)}, P_{k|k}^{(i)}, \theta_k^{(i)}\}_{i=1}^N$
\end{algorithmic}
\label{alg:RBSVGD}
\end{algorithm}

\section{Evaluation in Type 1 diabetes scenarios}

{We present two scenarios to evaluate the performance of the proposed \ac{RBSVGD} filter under both controlled and realistic conditions.
In the first scenario, we use the Hovorka model with time-varying parameters as the data-generating system. In this context, we want to assess performance in parameter tracking by leveraging ground-truth parameters to evaluate parameter estimation accuracy. In the second scenario, data generated by the \ac{OHSU} simulator are used to evaluate the proposed approach under more realistic and clinically relevant conditions.}


\subsection*{Performance metrics}
We evaluate the performance of the proposed RBSVGD filter  using 
the discrete-time Hovorka model~\cite{Hovorka2004}  as the model for the filter. 
The RBSVGD filter is compared against both a standard Extended Kalman Filter (EKF) and an augmented EKF (AEKF).  Filter performance is assessed using two metrics:  the 
Continuous Ranked Probability Score (CRPS) and the Root Mean Square Error (RMSE). 
The CRPS is a metric used to evaluate probabilistic forecasts by measuring the difference between the predicted cumulative distribution function (CDF) and the empirical CDF of the actual observation~\cite{gneiting2007strictly}.  
It measures how close a predicted cumulative distribution $F(x)$ is to the actual observed value $\bm{y}$, computed as the integrated squared difference between $F(x)$ and observed $\bm{y}$ as follows:
\begin{equation} \label{eq:crps}
   \mathrm{CRPS}\bigl(F,\, \bm{y}\bigr) = \int_{-\infty}^{\infty} \bigl(F(x)-\mathbf{1}_{\{x \geq \bm{y}\}}\bigr)^2 dx,
\end{equation}
with $\mathbf{1}_{\{x \geq \bm{y}\}}$ being an indicator function (1 if $\{x \geq \bm{y}\}$, else 0).
For the RBSVGD filter, whose state posterior is approximated as a Gaussian mixture 
(cf.~\eqref{eq:approx_posterior}), and for the EKF-based methods with Gaussian posteriors,
the CRPS in \eqref{eq:crps} is computed using the closed-form 
expression~\cite{grimit2006crps}. The RMSE for the 
blood glucose state over a horizon of $T$ steps is defined as
\begin{equation}
    \mathrm{RMSE} = \sqrt{\frac{1}{T}\sum_{k=1}^{T}\bigl(\bm{y}_k - \hat{\bm{y}}_k\bigr)^2},
    \label{eq:rmse}
\end{equation}
where $\bm{y}_k$ and $\hat{\bm{y}}_k$ denote the true and estimated blood glucose 
concentrations at time $k$, respectively. 
Lower values of both CRPS and RMSE
indicate better estimation performance. All reported CRPS and RMSE values for 
 blood glucose  $\bm{y}$ are expressed in mg/dL.

\subsection{Scenario I: The Hovorka Model as the data-generating system}

In this scenario, the data are generated using the Hovorka model~\eqref{eq:system}, considering three
time-varying physiological parameters $\theta_k = \left[k_e, S_{I,1}, S_{I,2}  \right]^{\top} \in \R^{3}$, whose 
trajectories are designed to capture a representative circadian and postprandial 
variations. These parameters are selected based on the comprehensive sensitivity analysis of the Hovorka model in~\cite{ESCORIHUELAALTABA_EDT_2025}, which ranked the influential parameters as those driving the largest variations in plasma glucose dynamics.
A single virtual patient with body 
weight $70\,\mathrm{kg}$ is simulated over a 24-hour horizon at a 
sampling interval of $T_s = 5\,\mathrm{min}$. Insulin delivery follows the 
 basal-bolus strategy of the McGill simulator~\cite{haidar2013stochastic}, in which the 
steady-state basal rate is computed analytically via inversion of the Hovorka 
model at a target fasting glucose level, and meal boluses are administered as 
preprandial corrections,  to maintain glucose within safe physiological 
limits. Blood-glucose $\bm{y}$  measurements are corrupted by zero-mean white Gaussian CGM sensor noise with a standard deviation of approximately   $5\,\mathrm{mg/dL}$.

The performance of the proposed RBSVGD filter is compared against: (i) a standard EKF with fixed nominal model parameters, and (ii) an AEKF that jointly 
estimates the states and a subset of selected parameters $\theta_{k}$. All filters are initialized with a common prior, and identical   process and measurement noise covariances $Q,R$ are chosen for all three filters. 
The hyperparameters of both the AEKF and the RBSVGD filter are selected via empirical tuning. In particular, the RBSVGD filter employs $N = 10$ particles and the \emph{Adam} optimizer with a step size of $\epsilon = 4 \times 10^{-5}$. 
For the AEKF, the parameter noise covariance is heuristically tuned to balance tracking responsiveness and estimation stability.

\begin{figure}[t]
    \centering
\includegraphics[width=\columnwidth]{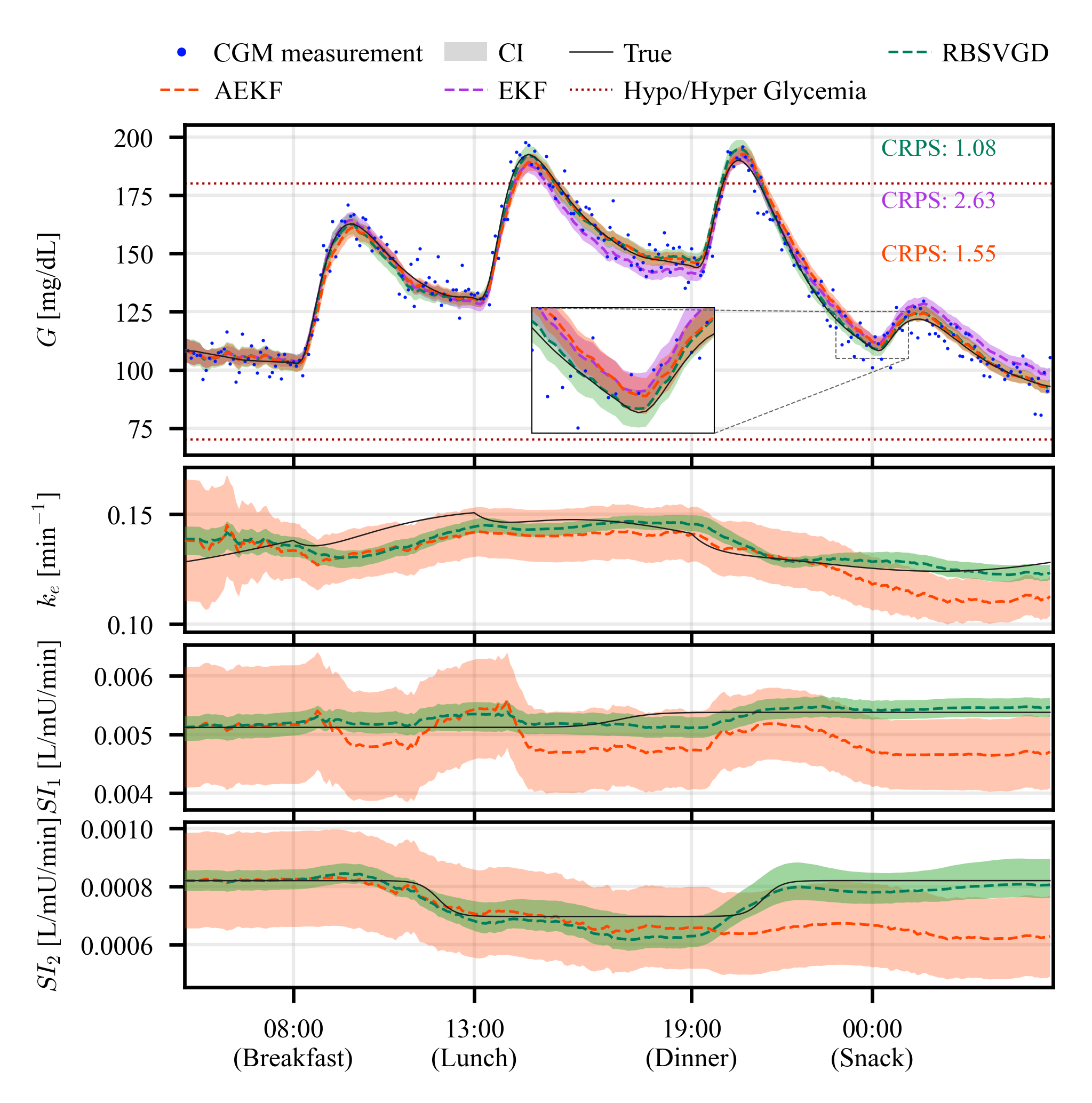}
    \caption{Scenario I:  \textit{Top panel:} True (black) \emph{vs} estimated by 
    EKF (dashed purple), AEKF (dashed orange) and RBSVGD (dashed green) filters. Noisy CGM measurements are indicated with blue dots. Shaded regions represent 95\% 
    confidence intervals.
    \textit{Rows 2-4:} Time-varying parameters 
    $k_e$, $S_{I,1}$, and $S_{I,2}$:  True (black) \emph{vs} estimated mean trajectories with AEKF (orange dashed) and RBSVGD (green dashed). The shaded orange region is the $\pm 2$ standard deviation of the AEKF Gaussian parameter posterior. The shaded green region shows the area between all particle trajectories of RBSVGD.   
  }
    \label{fig:scenario1}
\end{figure}

Figure~\ref{fig:scenario1} illustrates the blood glucose estimation and the 
 tracking of the parameters   $k_e$, $S_{I,1}$, and $S_{I,2}$ over a 24-hour period using the EKF, AEKF and RBSVGD filters. The results show  that the proposed RBSVGD filter achieves  more accurate tracking of the true blood glucose compared to both the EKF and AEKF. Moreover, it provides a tighter and better-calibrated confidence intervals 
around the true glucose trajectory. This improvement is reflected  
in the  lower CRPS  value of $1.08$ obtained  by RBSVGD, compared to $1.55$ for the AEKF and $2.63$ for the EKF. Furthermore, the rows  $2-4$ in Fig.~\ref{fig:scenario1} depict the tracking performance for the time-varying parameters $k_e$, $S_{I,1}$, and $S_{I,2}$. The results indicate that
the RBSVGD 
filter tracks the time-varying parameter trajectories more accurately, owing to its non-parametric representation of the parameter posterior, which is not restricted to a Gaussian approximation as in the AEKF.

This experiment confirms the RBSVGD filter's ability to track the state and uncertain parameters more effectively.


\subsection{Scenario   II: OHSU Simulator with 20 Virtual Patients}

In the second scenario, data are generated using the \ac{OHSU} Type~1 Diabetes simulator~\cite{OHSU_simulator_2019}, providing a more realistic evaluation setting. The \ac{OHSU} has been used as a preclinical step to validate diabetes technology prior to clinical testing. This simulator extends the modified Hovorka glucose-insulin model with an explicit aerobic exercise module and glucagon secretion. It enables the simulation of the effects of physical activity on glucose dynamics, including increased peripheral glucose uptake in active muscle tissue, enhanced peripheral insulin uptake, and modulation of hepatic glucose production~\cite{OHSU_simulator_2019}. A stochastic virtual patient population is generated by sampling from a distribution of possible insulin sensitivities, by capturing inter-patient variability. The exercise model, adapted from~\cite{HERNANDEZORDONEZ2008744}, modulates the insulin sensitivity factors during exercise bouts based on intensity represented as percentage of maximum oxygen consumption (PVO$_{\mathrm{2max}}$) and fraction of active muscular mass (PAMM).

The experimental setup for the in silico evaluation involved $N_p = 20$ virtual patients randomly selected from the simulator population. A two-day simulation scenario, with three meals per day is scheduled as follows: on Day 1 at 08:00 (breakfast) $70\,\mathrm{g}$, 12:30 (lunch) $100\,\mathrm{g}$, and 20:00 (dinner) $100\,\mathrm{g}$; on Day 2 at 08:30 (breakfast) $35\,\mathrm{g}$, 12:50 (lunch) $79\,\mathrm{g}$, and 19:00 (dinner) $117\,\mathrm{g}$. On the first day only, a single aerobic exercise bout of 120 minutes was performed, starting 3.5 hours after lunch. The simulation was configured with PAMM having 50\% and PVO$_{\mathrm{2max}}$ with 60\%, consistent with moderate-intensity aerobic activity used in prior model validations. The exercise response was modeled using the simulator's built-in aerobic module, which dynamically adjusts peripheral insulin uptake (PIU), peripheral glucose uptake (PGU), and hepatic glucose production (HGP) during the bout. 

The simulator provided a multiple daily injection (MDI) therapy consisting of constant basal insulin and bolus to compensate for meals. The hypoglycemia treatment module followed the standard 15-15 rule, triggering the administration of 15 g of rescue carbohydrates whenever measured glucose values fell below $70\,\mathrm{mg/dL}$. An additional $15\,\mathrm{g}$ dose is administered every 15 minutes until glucose recovers more than $70\,\mathrm{mg/dL}$. Following each rescue carbohydrate administration, the basal insulin infusion rate is reduced to 25\% of the nominal value and maintained for 40 minutes to prevent hypoglycemia recurrence.

To account for the exercise response, we consider $\theta_k = \left[S_{I,1}, S_{I,2}, S_{I,3}  \right]^{\top} \in \R^{3}$ the time-varying parameters, as these parameters are increased during the exercise bout as functions of PIU, PGU and HGP~\cite{OHSU_simulator_2019}. 
 The performance of the proposed RBSVGD filter is compared against a standard EKF with fixed nominal model parameters and an AEKF. 
The filter hyperparameters are determined through empirical tuning. 
In particular, the RBSVGD filter employs $N = 5$ particles and the \emph{Adam} optimizer 
with a step size of $\epsilon = 1 \times 10^{-4}$. All other 
hyperparameters, including the noise covariances, are kept identical to those used in Scenario I.

\begin{figure}[t]
    \centering
    \includegraphics[width=\columnwidth]{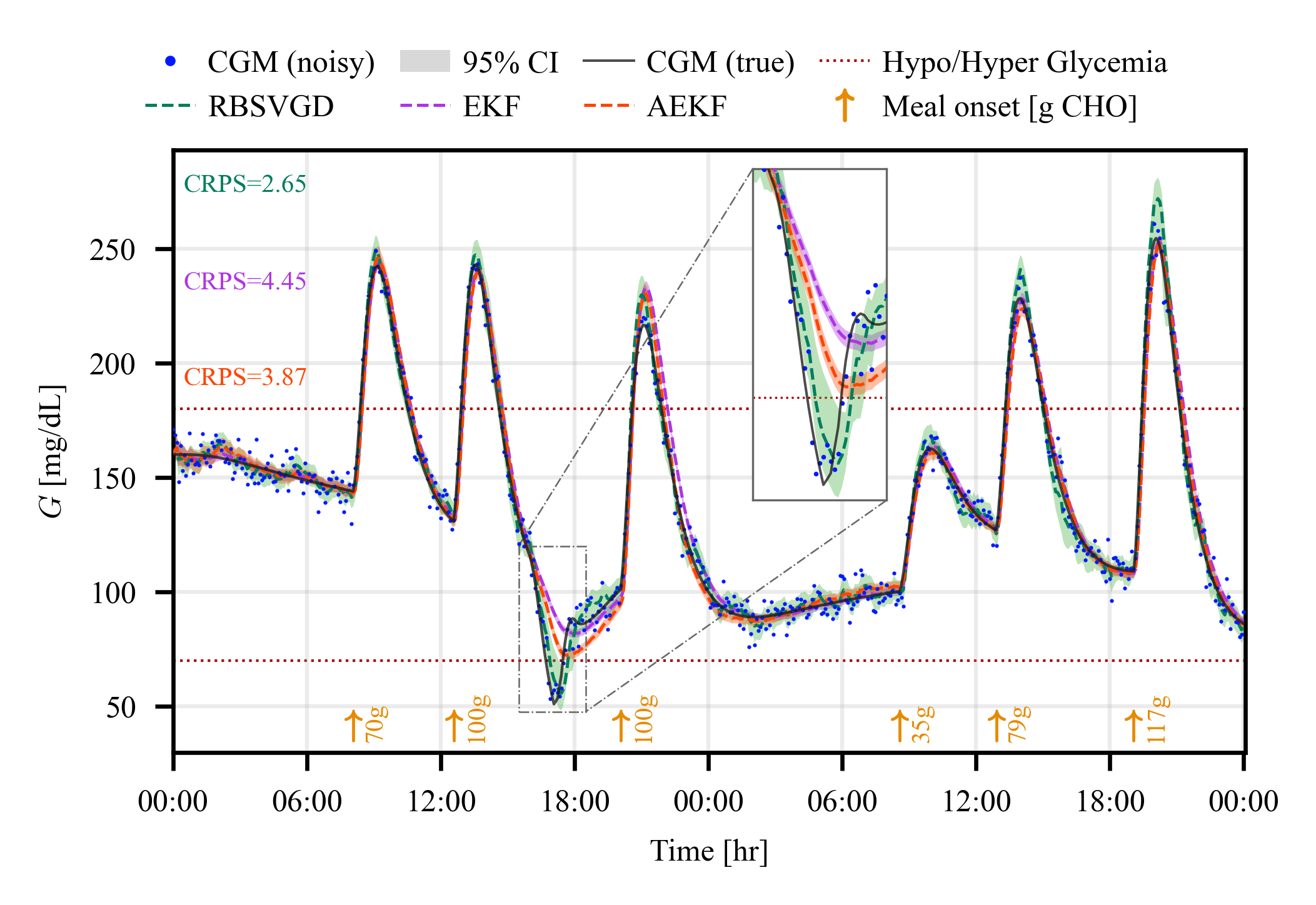}
    \caption{Scenario II: True (black) \emph{vs} estimated blood glucose  for a representative virtual 
    patient from the OHSU simulator cohort with EKF (dashed purple), AEKF (dashed orange) and RBSVGD (dashed green) filters. The blue dots indicate noisy CGM measurements. Shaded regions represent 95\% 
    confidence intervals. The zoomed window corresponds to exercise period of $120$ minutes. 
    }
\label{fig:scenario2_glucose}
\end{figure}

Figure~\ref{fig:scenario2_glucose} presents representative blood glucose estimation 
results for a single patient from the cohort. The RBSVGD filter yields estimates 
that more closely track the true glucose trajectory, particularly during the exercise period as highlighted with the box in Fig.~\ref{fig:scenario2_glucose},  while maintaining well-calibrated 
uncertainty bounds, demonstrating robustness to the more complex, 
simulator-generated dynamics. This improvement is further reflected in the CRPS score of $2.65$ achieved by RBSVGD, which is lower compared to $3.87$ for the AEKF and  $4.45$ obtained with the EKF.  

\begin{figure}[t]
    \centering
    \includegraphics[width=\columnwidth]{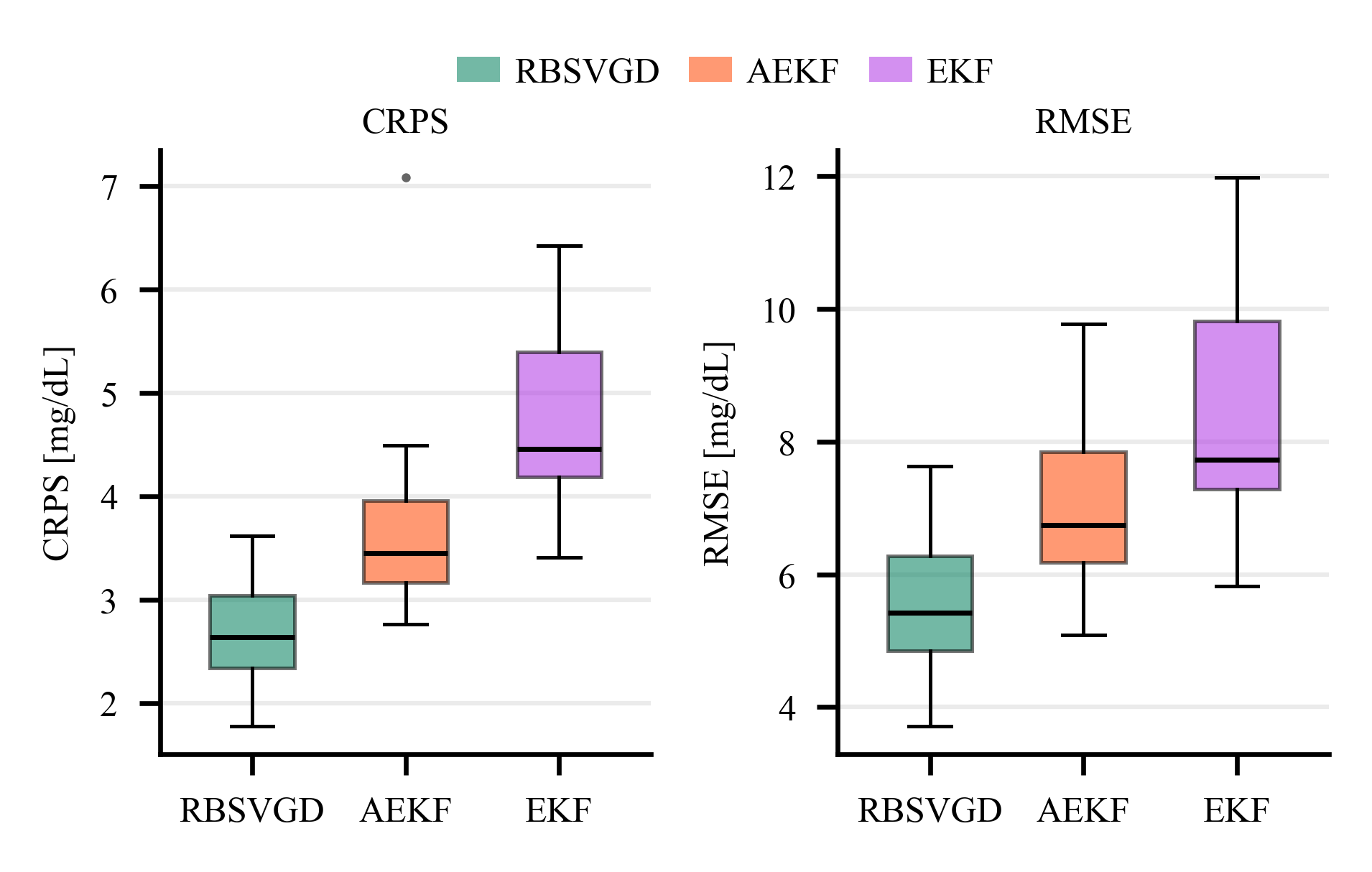}
    \caption{Scenario II: Box plots of the CRPS [mg/dL] (\textit{left panel}) and RMSE 
    [mg/dL] (\textit{right panel}) for blood glucose estimation across $N_p = 20$ 
    virtual patients from the OHSU simulator. Results are shown for the EKF 
    (fixed parameters), AEKF, and the proposed RBSVGD filter, computed for $48$-hour simulation horizon.} 
\label{fig:scenario2_boxplot}
\end{figure}

To assess the robustness of the proposed approach, the filters are applied to state estimation across a cohort of 20 virtual patients. Fig.~\ref{fig:scenario2_boxplot} presents box plots of the CRPS and RMSE obtained for the EKF, AEKF, and the proposed RBSVGD filter across 
all $N_p = 20$  patients. 
The incorporation of parameter-varying Hovorka model with time-varying parameters $\theta_k$ in the AEKF and RBSVGD filters enables them to account for both  intra- and inter-patient variability, resulting in  improved performance compared to the fixed parameter  EKF as seen in Fig.~\ref{fig:scenario2_boxplot}. Notably, the RBSVGD filter achieves consistently lower CRPS and RMSE values across 
the patient cohort than AEKF, with reduced inter-patient variability. 

Finally, to further investigate the impact of time-varying parameters, 
Table~\ref{tab:scenario2_day} reports the mean and standard deviation of the CRPS and RMSE values, 
disaggregated by day across the $N_p = 20$ virtual patients. 
This analysis highlights the effect of exercise, which is present on Day~1 and absent on Day~2. Accordingly, the RBSVGD and AEKF filters are configured with time-varying parameters on Day~1 to capture exercise-induced variability, while parameters are held fixed on Day~2.  
\begin{table}[tb!]
    \centering
    \caption{Scenario II: Mean~$\pm$~standard deviation of CRPS [mg/dL] and 
    RMSE [mg/dL] for blood glucose estimation across $N_p = 20$ virtual 
    patients, reported separately for Day~1 and Day~2 of the 48-hour 
    simulation horizon.}
\label{tab:scenario2_day}
    \begin{tabular}{lccc}
        \toprule
        & RBSVGD & AEKF & EKF \\
        \midrule
        \multicolumn{4}{l}{\textit{CRPS [mg/dL]}} \\
        \quad Day 1 
            & $3.52 \pm 0.81$ & $5.02 \pm 1.01$ & $6.06 \pm 1.44$ \\
        \quad Day 2 
            & $1.84 \pm 0.37$ & $2.44 \pm 1.12$ & $3.37 \pm 0.50$ \\
        \midrule
        \multicolumn{4}{l}{\textit{RMSE [mg/dL]}} \\
        \quad Day 1 
            & $6.86 \pm 1.58$ & $8.80 \pm 1.69$ & $10.29 \pm 2.45$ \\
        \quad Day 2 
            & $3.97 \pm 0.70$ & $4.72 \pm 1.03$ & $5.98 \pm 0.70$ \\
        \bottomrule
    \end{tabular}
\end{table}
The RBSVGD filter outperforms both benchmark methods on each day, with the 
performance gap being most pronounced on Day~1, when the parameters 
$S_{I,1}$, $S_{I,2}$, and $S_{I,3}$ are considered to be time-varying. On Day~2, where 
the parameters are held constant, differences across filters are reduced. This observation suggests that the primary advantage of the RBSVGD filter stems from 
its ability to effectively incorporate parameter-varying Hovorka model, with  time-varying parameters estimated through its general non-parametric 
posterior representation.

\section{Conclusions}
This paper presented a \ac{RBSVGD} filter for online joint state and parameter estimation in nonlinear time-varying physiological models for type 1 diabetes. Compared with a standard \ac{EKF} with fixed parameters and an \ac{AEKF} for joint parameter estimation, the RBSVGD approach demonstrated  superior performance. It achieved higher accuracy in glucose prediction 
while providing reliable tracking of time-varying parameters, particularly under dynamic insulin sensitivity conditions. The main advantage arises from the filter’s ability to handle non-Gaussian uncertainties and parameter variability through a non-parametric posterior representation.
These results highlight the potential of advanced filtering techniques to overcome the limitations of classical \ac{EKF} methods in free-living conditions. 
Future work will focus on evaluating performance under other continuous variations such as circadian rhythms, and on real-world patient data and for glucose-predictions.
\appendix
\label{sec:Hovorka}

The continuous--time Hovorka Glucose--Insulin model dynamics \eqref{eq:system} is defined as: 
\begin{subequations}
\label{eq:hovorka_odes}
\begin{align}
  \ddt{Q_1(t)} &= \frac{D_2(t)}{\tau_D} - x_1(t)\,Q_1(t) - F_{01}^{c}(t)\notag\\
               &\quad - F_R(t) + k_{12}\,Q_2(t) + \mathrm{EGP}(t),
               \label{eq:dQ1}\\[4pt]
  \ddt{Q_2(t)} &= x_1(t)\,Q_1(t) - \bigl(k_{12} + x_2(t)\bigr)\,Q_2(t),
               \label{eq:dQ2}\\[4pt]
  \ddt{S_1(t)} &= u_1(t) - \frac{S_1(t)}{\tau_S},
               \label{eq:dS1}\\[4pt]
  \ddt{S_2(t)} &= \frac{S_1(t)}{\tau_S} - \frac{S_2(t)}{\tau_S},
               \label{eq:dS2}\\[4pt]
  \ddt{I(t)}   &= \frac{S_2(t)}{\tau_S\,V_I} - k_e(t)\,I(t),
               \label{eq:dI}\\[4pt]
  \ddt{x_1(t)} &= -k_{a_{1}}\,x_1(t) + k_{b_{1}}\,I(t),
               \label{eq:dx1}\\[4pt]
  \ddt{x_2(t)} &= -k_{a_{2}}\,x_2(t) + k_{b_{2}}\,I(t),
               \label{eq:dx2}\\[4pt]
  \ddt{x_3(t)} &= -k_{a_{3}}\,x_3(t) + k_{b_{3}}\,I(t),
               \label{eq:dx3}\\[4pt]
  \ddt{D_1(t)} &= A_G\,\frac{u_2(t)}{180.16\times 10^{3}} - \frac{D_1(t)}{\tau_D},
               \label{eq:dD1}\\[4pt]
  \ddt{D_2(t)} &= \frac{D_1(t)}{\tau_D} - \frac{D_2(t)}{\tau_D},
               \label{eq:dD2}\\[4pt]
  G(t) &= \frac{Q_1(t)}{V_G}.
               \label{eq:G}
\end{align}
\end{subequations}
\noindent
The auxiliary variables appearing in~\eqref{eq:dQ1} are defined as:
\begin{align*}
  F_{01}^{c}(t) &=
  \begin{cases}
    F_{01},                        & G(t) \geq \SI{4.5}{\milli\mol\per\liter},\\[2pt]
    \dfrac{F_{01}\,G(t)}{4.5},   & G(t) < \SI{4.5}{\milli\mol\per\liter},
  \end{cases}\\[4pt]
  F_R(t) &= \max{(0, 0.003\!\left(G(t) - 9.0\right)V_G)}\\[4pt]
  \mathrm{EGP}(t) &= \max{(0, \mathrm{EGP}_0\,\bigl(1 - x_3(t)\bigr)).}
\end{align*}
The detailed description of the states and parameters are given in Section~\ref{sec:problem_set}.

\bibliographystyle{plain}
\bibliography{references}

\end{document}